\documentclass[conference]{IEEEtran}
\IEEEoverridecommandlockouts
\usepackage{cite}
\usepackage{amsmath,amssymb,amsfonts}
\usepackage{array,multirow,graphicx}
\usepackage{pgfplots}
\usepackage[T1]{fontenc}
\pgfplotsset{compat=1.10}
\usepackage{tikz}
\usepackage{array,booktabs}
\usepackage{makecell, pict2e}
\usetikzlibrary{patterns}
\pgfplotsset{width=16cm,compat=1.8}
\usepgfplotslibrary{groupplots}
\usepackage{pgfplots}
\usepgfplotslibrary{groupplots}
\pgfplotsset{compat=1.12}
\usetikzlibrary{patterns}
\usepackage{algorithmic}
\usepackage{diagbox}
\usepackage{graphicx}
\usepackage{textcomp}
\usepackage{xcolor}
\usepackage{lipsum,tabularx}
\usepackage{booktabs,graphicx}
\usepackage{amsmath}
\usepackage{caption,subcaption}
\usepackage{hyperref}
\usepackage{rotating}
\usepackage{array,makecell,multirow}
\usepackage{pgfplotstable}
\pgfplotsset{compat=1.11,
        /pgfplots/ybar legend/.style={
        /pgfplots/legend image code/.code={%
        \draw[##1,/tikz/.cd,bar width=3pt,yshift=-0.2em,bar shift=0pt]
                plot coordinates {(0cm,0.8em)};},
},
}

\def\BibTeX{{\rm B\kern-.05em{\sc i\kern-.025em b}\kern-.08em
    T\kern-.1667em\lower.7ex\hbox{E}\kern-.125emX}}
\newcolumntype{P}[1]{>{\centering\arraybackslash}p{#1}}
\begin{document}
\title{Design Smell Analysis for Developing and Established Open Source Java Software\\
}
\author{\IEEEauthorblockN{Asif Imran}
\IEEEauthorblockA{\textit{IEEE Student Member}\\
\textit{Department of CSE, }\\
\textit{University at Buffalo (SUNY)}\\
New York, USA \\
asifimra@buffalo.edu}
\and
\IEEEauthorblockN{Tevfik Kosar}
\IEEEauthorblockA{\textit{IEEE Member}\\
\textit{Department of CSE, }\\
\textit{University at Buffalo (SUNY)}\\
New York, USA \\
tkosar@buffalo.edu}
}

\maketitle

\begin{abstract}
Software design smells are design attributes which violate the fundamental design principles. Design smells are a key cause of design debt. Although the activities of design smell identification and measurement are predominantly considered in current literature, those which identify and communicate which design smells occur more frequently in newly developing software and which ones are more dominant in established software have been studied to a limited extent. This research describes a mechanism for identifying the design smells that are more prevalent in developing and established software respectively. A tool is provided which is used for design smell detection by analyzing large volume of source code. More specifically, 164,609 Lines of Code (LoC) and 5,712 class files of six developing and 244,930 LoC and 12,048 class files of five established open source Java software are analyzed. Obtained results show that out of the 4,020 occurrences of smells which were made for nine pre-selected types of design smells, 1,643 design smells were detected for developing software, which mainly consisted of four specific types of smells. For established software, 2,397 design smells were observed which predominantly consisted of four other types of smells. The remaining design smell was equally prevalent in both developing and established software. Desirable precision values ranging from 72.9\% to 84.1\% were obtained for the tool. 

\end{abstract}


\section{Introduction}
\label{introduction}

Software design engineering is an important activity which requires careful application of design guidelines. Design issues contribute to 64\% of software defects as a study by Jones et al. \cite{softqual} highlighted. Hence, quality and maintainability of software are significantly affected by design problems. One of the important design issues is a software suffering from design smells. Design smells are design aspects which violate the fundamental design principles and negatively affect the design of the software \cite{suryanarayana2014refactoring}. A software which has large volume of design smells contributes to design debt which in-turn increases technical debt. In recent years, a greater emphasis to reduce design smells has been given by software companies, engineers and researchers \cite{fontana2016comparing}. We assume that despite the increased importance, both developing and established software suffer from diverse design smells, which is undesirable and hampers long term sustainability. Research efforts are needed to identify which smells are more prevalent in developing software and which occur more frequently in established software. The advantage of specifying a subset of design smells from a large set will enable software engineers to focus on refactoring those and not spending time and refactoring effort on other smells, which will help to reduce design debt. However, no study has been conducted which analyses the frequency of occurrence of various design smells in developing software and established software.


List of 25 design smells were provided by Suryanarayana et al. \cite{suryanarayana2014refactoring} which focused on fundamental design aspects. Although this list is useful, it does not specify which smells occur more frequently in developing software and which ones occur more significantly in established software. A design debt prioritization using a portfolio matrix was provided by Plosch et al. \cite{ plosch2018design} and communication of remedy actions were mentioned. However, which smells must be addressed with a higher priority in developing software and which ones can be ignored was not elaborated. A model for detecting cyclic dependency and hub-like dependency smells using link prediction techniques was discussed by Diaz-Pace \cite{diaz2018towards}. However, detection of more impactful design smells like \textit{Unutilized Abstraction} \cite{suryanarayana2014refactoring} which tend to occur to a greater volume has not been addressed. A catalogue to list all architectural smell detection tools together with their operating platforms was provided by Azadi et al. \cite{azadi2019architectural}. However, the catalogue does not provide a comparative analysis as to whether the tools can successfully detect architectural smells occurring in developing and established software respectively.

Based on the above motivation, this paper contributes primarily to determining whether a certain design smell occurs more frequently in developing or established software. The assumption of developing and established software within the scope of this paper are provided later. First, the paper aims to detect software properties. Next, the tool for design smell detection based on pseudo-model generation using \textit{Abstract Syntax Tree (AST)} \cite{neamtiu2005understanding} is used to analyze 164,609 LoC and 5,712 class files of six developing and 244,930 LoC and 12,048 class files of five established open source Java software. Afterwards, causal relationship between the software properties and the design smells are set up. Finally, precision and recall are calculated to identify the performance of the tool. 

Analysis of results show  desirable output regarding occurrence of a specific smell in greater volume in either developing or established software. This paper focuses on nine design smells due to the high percentage of occurrence of those in the analyzed software. We found that design smells namely \textit{Broken Hierarchy, Deficient Encapsulation, Missing Hierarchy} and \textit{Wide Hierarchy} contribute more to design debt of developing software since those occur in significantly greater percentage in developing software. On the other hand, design smells like \textit{Insufficient Modularization, Cyclic-Dependent Modularization, Unnecessary Abstraction} and \textit{Multifaceted Abstraction} are seen to occur in a greater volume in established software, thus contributing more towards the technical debt of those. The smell \textit{Unutilized Abstraction} contribute to the design debt of both developing and established software stacks. A total of 4,020 occurrences could be observed. Quantitative values of the frequency of occurrence for all the smells is provided in the results analysis section. The results are desirable as precision values ranging from \textit{72.9\%} to \textit{84.1\%} are obtained for the analyzed software.

Based on the above information, the major contributions of this paper can be stated as follows:-
\begin{itemize}
    \item Provide a tool for design smell detection in open source Java software and analyze its performance via calculation of precision.
    \item Identify which design smells are more prevalent in developing software and which ones are dominant in established software.
    \item Establish relationship between key software quality properties and occurrence of design smells.
\end{itemize}

The rest of the paper proceeds as follows. Section \ref{background} provides a summary of the background information regarding Open Source Software (OSS) and Design Smells. Section \ref{empirical} illustrates the research questions, identifies the selected software systems for this study and describes the implementation of the tool for design smell detection and recording. Section \ref{analysis} provides the obtained results and analyses those. Section \ref{relatedwork} discusses the related work, and Section \ref{conclusion} concludes the paper and discusses future research directions.

\section{Background}
\label{background}
This section identifies the background information to understand the scope and context of this study. It provides information regarding \textit{Open Source Software (OSS)} and design smells detection. 
\subsection{Open Source Software (OSS)}

\subsubsection{Definition and popularity} Software which allow the users to obtain, modify, use, run and improve it free of cost is called an $OSS$ \cite{bessen2002good}. There is a large number of $OSS$ which can be obtained through websites like $BitBucket, Github, Launchpad, SourceForge$ etc \cite{15best}. Those software have been gaining increasing popularity both in the private and public sectors. Java is a popular language for writing large scale $OSS$ and till date a significant portion of $OSS$ have been written with this language. It provides state of the art development toolkits and provides useful frameworks to solve complex problems. According to \textit{TIOBE Software's} latest Programming Community Index, Java has the largest number of server side implementations in projects, covering a total of 15.978\% \cite{Grechanik:2010:EIL:1852786.1852801}. This research bases the analysis on open source Java projects. 

\subsubsection{OSS Design} When a company open sources its software, then it is called an open source system \cite{5631694}. Initially those software may have limited number of developers. However, as it gains popularity, a significant number of contributors join and leave. Some contributors focus on fixing the bugs and solving errors, others focus on adding new features. Many new developers focus on learning to design and develop better software by contributing to open source projects. The developers use collaboration tools to maintain various versions of the open source systems. We analyzed a total of 409,539 LoC of open source developing and established systems in Java. Given the popularity of open source systems among developers and companies, it is important to detect the design smells of both developing and established open source systems \cite{plosch2018design}. 
\subsection{Design Smells}
Design smells are considered to be design aspects which violate the fundamental design principles and negatively affect the design of the software \cite{suryanarayana2014refactoring}. This paper determine whether a design smell occurs more frequently in developing software or is it more frequent in established software. This information could guide software engineers to identify and focus on specific smells which occur in their systems. Table \ref{designsmellstable} describes the top nine design smells out of twenty five smells which are taken in consideration due to the high occurrence.

\begin{table}
    \centering
    \scalebox{1}{
    \begin{tabular}{|p{0.4\textwidth}|}
        \hline \textbf{Unutilized Abstraction}: This design smell occurs when an abstraction is declared, however its implementation is missing in the code \cite{suryanarayana2014refactoring}. For example, a class \textit{ABC} may be declared by the software developer, however if it is not used anywhere in the code, then it is a \textit{Unutilized Abstraction} design smell.\\
        \hline\hline
        \textbf{Insufficient Modularization}: When an abstraction is large and complicated, providing a scope of further modularization, it is said to suffer from this design smell \cite{suryanarayana2014refactoring}. This smell occurs when the class is large in size, multiple class definitions are present in a file or the complexity of the class is high.\\
        \hline\hline
        \textbf{Broken Hierarchy}: This occurs when the IS-A linkage between a supertype and subtype is absent. This interferes with the substitution capability among the two classes \cite{suryanarayana2014refactoring}.\\
        \hline\hline
        \textbf{Deficient Encapsulation}: When an abstraction given more accessibility to the users than it is required, it threatens the security of the software. The presence of this smell is called \textit{Deficient Encapsulation} \cite{suryanarayana2014refactoring}.\\
        \hline\hline
        \textbf{Cyclic Dependent Modularization:} This smell occurs when the software violates the acyclic modularization technique \cite{sarkar2012measuring}. It occurs when one abstraction is cyclically dependent on another. If this design smell persists, then a change in one abstraction can have a ripple effect on other abstractions in the software \cite{oyetoyan2013study}. \\
        \hline\hline
        \textbf{Unnecessary Abstraction:} When a software has multiple layers of abstraction and many of the intermediate layers are not required, the software is said to suffer from \textit{Unnecessary Abstractions} \cite{unnecessary}.\\
        \hline\hline
        \textbf{Wide Hierarchy}: This smell is prevalent when an inheritance tree has a wide breadth and the intermediate types are missing \cite{suryanarayana2014refactoring}.\\ 
        \hline\hline
        \textbf{Imperative Abstraction}: It occurs when a class has limited functionality within an operation \cite{suryanarayana2014refactoring}. It may also happen when an operation is converted to a class.\\ 
        \hline\hline
        \textbf{Multifaceted Abstraction}: When an abstraction is responsible for multiple functionalities, managing it may become troublesome, resulting in this design smell \cite{suryanarayana2014refactoring}. Also, many functionalities might be affected if an error occurs in the abstraction.\\
        \hline
    \end{tabular}}
    \caption{Description of the selected design smells}
    \label{designsmellstable}
\end{table}

\begin{figure}
    \centering
    \includegraphics[width=6.8cm, height=8cm]{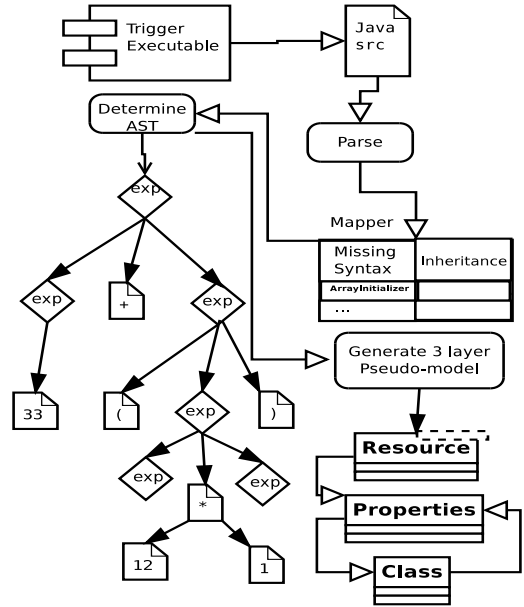}
    \caption{Representation for generating AST and pseudomodel for design smell analysis}
    \label{fig:framework}
\end{figure}
\section{Empirical study setup}
\label{empirical}
This study focuses on identifying which design smells occur more frequently during the development and matured phases of Java software. The following subsections identify the research questions and data collection strategy.
\subsection{Research Questions}
The following research questions are addressed in this paper.\\ 
\textbf{RQ1: To what extent is the detection of design smells in developing and established software accurate?}
\begin{itemize}
    \item \textbf{RQ1.1:} What is the precision of the provided approach?
    \item \textbf{RQ1.2:} What value of recall is achieved by the design smell detection scheme? 
\end{itemize}
Software evolves from initial to matured phases, it may undergo significant changes in design. Hence, the tool should detect design smells present in both developing and established software. Via identification of precision and recall, we aim to establish the performance of the tool.
\textbf{RQ2: Are all design smells equally prevalent in developing and established open source software?}
\begin{itemize}
    \item \textbf{RQ2.1: }Which design smells occur more frequently when a new software is developed? 
    \item \textbf{RQ2.2: }Which design smells occur more frequently after a software has served the users for significant time and has become established?
\end{itemize}
This paper provided a tool to detect design related smells which occur in new and established software. It grouped the smells to either belong to developing or established software based on the frequency of occurrence. Additionally via analyzing the properties of multiple new and established software, the affect of the mentioned attributes on the occurrence of design smells in developing and established software are shown. The answer to those research questions can help software engineers focus on specific group of design smells rather than the entire lot.\\

\subsection{Selection of Software Systems}
\textit{Open Source Software (OSS)} divides in two stacks, the first stack represented six developing software collection and the second stack contained five established software. The criteria for determining developing and established software are provided in the following. 


\textbf{Developing Software:} In the scope of this paper, a software is considered to belong to the developing stack if it satisfies the following requirements:
\begin{itemize}
    \item Must have less than or equal to 2,000 commits
    \item Repository shall have less than or equal to 30 contributors
     \item Git repository is active and latest commit is within 9 months of writing this paper
    \item Should not have more than 2 releases so far
\end{itemize}
\textbf{Established Software:} We focused on specific criteria for selecting established Java software which are provided below. 
\begin{itemize}
    \item Must have more than 2,000 commits
    \item Have more than 30 contributors
    \item Atleast 2 or more releases
\end{itemize}

Table \ref{list} summarizes the list of eleven selected systems among which the developing projects are primarily Managed File Transfer Software (MFT) identified in \cite{imran2018onedatashare}. For each of the systems belonging to either the developing stack and established stack, we highlight the domain, life-time, $KLoc$, number of commits and number of contributors for those. We tested those software to detect the design smells using the tool described in the following subsection.

\subsection{Implementing Smell Detection Tool} The tool detects smells similar to \textit{Designite} \cite{7809461}. However, \textit{Designite} is designed for detection of smells for C\# whereas our tool is modified for detecting design smells in Java code. The output of detection are stored in a provenance file. Another difference is that \textit{Designite} \cite{7809461} detects both design and code smells in C\# code and has a user interface whereas we focus on detection of design smells for Java code and use a command line interface \textit{CLI}. Figure \ref{fig:framework} identifies the process of \textit{AST} generation for obtaining the design smells.  

Firstly, Java source code are taken as input then used to form an \textit{Abstract Syntax Tree (AST)}. The first phase includes lexical analysis followed by syntax analysis. To form the \textit{AST, Janett} \cite{janett} has been used which converts constructs and calls to Java libraries to nodes of $AST$. Hence the \textit{AST} can be formed using \textit{NRefactory} \cite{grunwald2012nrefactory} after parsing the Java code with \textit{Janett}. Special Java constructs like $Abstract classes$, \textit{ArrayInitializer}, etc which are not present are converted using a mapper. \textit{IKVM} library is used to translate syntax and constructs, using inheritance whenever is necessary. 

\begin{table}
\centering
 \begin{tabular}{|c| l r r r r|} 
 \cline{2-6}
 \multicolumn{1}{c|}{}&\scriptsize System & \scriptsize Activity &\scriptsize KLoC &\scriptsize \scriptsize Commits &\scriptsize Contrib.\\ [0.3ex] 
 \cline{1-6}
  \parbox[t]{2mm}{\multirow{6}{*}{\rotatebox[origin=c]{90}{Developing}}} 
 & \scriptsize Yade\cite{sisiaridis2016framework} &  \scriptsize 02/14-02/19 &\scriptsize 29.051 &\scriptsize 1871 &\scriptsize 22 \\ 
 &\scriptsize Divconq\cite{sisiaridis2016framework} &\scriptsize 09/14-02/19 &\scriptsize 64.448 &\scriptsize 88 &\scriptsize 2 \\
 &\scriptsize Mover.io\cite{ross2014managed} &\scriptsize 05/14-02/18 &\scriptsize 23.002 &\scriptsize 1644 &\scriptsize 20 \\
 &\scriptsize Waarp\cite{waarp2018} &\scriptsize 08/09-03/19 &\scriptsize 24.267 &\scriptsize 385 & \scriptsize 3 \\
 &\scriptsize OneDataShare\cite{imran2018onedatashare} &\scriptsize 11/18-05/19 &\scriptsize 17.201 &\scriptsize 412 &\scriptsize 10 \\
 &\scriptsize Accord\cite{accord2018} &\scriptsize 07/10-05/19 &\scriptsize 19.391 &\scriptsize 352 &\scriptsize 3 \\ 
 \hline
 \parbox[t]{2mm}{\multirow{5}{*}{\rotatebox[origin=c]{90}{Established}}} 
 &\scriptsize RxJava\cite{nurkiewicz2016reactive} &\scriptsize 03/12-04/19 &\scriptsize 141.038 &\scriptsize 5531 &\scriptsize 240 \\
 &\scriptsize JAMES\cite{james2018} &\scriptsize 09/06-03/19 &\scriptsize 27.271 &\scriptsize 868 &\scriptsize 86 \\
&\scriptsize Zimbra\cite{zimbra} &\scriptsize 08/05-03/19 &\scriptsize 24.698 &\scriptsize 15052 &\scriptsize 68 \\
&\scriptsize ApacheCommons\cite{commons2013math} &\scriptsize 07/02-03/19 &\scriptsize 49.534 &\scriptsize 5446 &\scriptsize 115 \\
&\scriptsize LoboEvolution \cite{cikryt2015evaluating} &\scriptsize 10/14-11/18 &\scriptsize 17.331 &\scriptsize 788 &\scriptsize 42 \\
 \hline
\end{tabular}
\caption{List of selected systems for the study}
\label{list}
\end{table}

We have a simple version of the model from the $AST$ consisting of three layers. The lowest layer consists of the data elements and objects in Java code. The second layer is a description of the elements found in the lowest layer. The third layer consists of the data types and namespaces of the objects. Using the information captured in the pseudo-model, design smells in the Java code are detected with pre-specified rules which are consistent for both developing and established software. Finally, those smells are written in a provenance file which can be used for analysis.

To use the system the \textit{CLI} will enable running the tool to obtain frequency of specific design smells for both developing and established software. This is followed by passing the path of the Java source files of the target software. Next, the parameter and threshold values are passed. Afterwards, the design smells can be detected. Once detected, the tool writes the output of detection to a $provenance.log$ file which can be later further analyzed considering it as a structured data-set of smells to obtain frequency of design smell prevalent in the software. 

\section{Data analysis and results}
This section analyses the results which are obtained by applying the tool to the selected developing and established list of software. The affect of various software properties on design smell of developing and established software is provided and the diversity on the occurrence of nine target design smells are studied to define the frequency of those. 
\label{analysis}
\subsection{Analysis of Software Quality Properties}
\subsubsection{Child class}
\begin{figure*}
        \centering
        \begin{tikzpicture}
            \begin{groupplot}[
                    legend columns=-1,
                    legend entries={{Mover.io},{Accord},{Waarp},{OneDataShare},{Yade},{Divconq},{RxJava},{Zimbra},{ApacheCommons},{James},{LoboEvolution}},
                    legend to name=CombinedLegendBar,
                    footnotesize,
                    ybar legend,
                    width = 6.8cm,
                    height = 4.3cm,
                    ylabel=percentage(\%),
                    xlabel style={yshift=-1cm},
                    group style={
                    group size=3 by 1,
                    xlabels at=edge bottom,
                    ylabels at=edge left,
                    }]
                \nextgroupplot[title={\scriptsize (a) Percentage of child classes to total classes}, xticklabels=\empty]
                    \addplot[ybar, pattern=horizontal lines, pattern color=red] coordinates {  (1, 7.26)};
                    \addplot[ybar, pattern=vertical lines, pattern color=red] coordinates { (2, 9.07)};
                    \addplot[ybar, pattern=grid, pattern color=red] coordinates {  (3, 8.17)};
                    \addplot[ybar, pattern=dots, pattern color=red] coordinates {  (4, 4.66)};
                    \addplot[ybar, pattern=north east lines, pattern color=red] coordinates {  (5, 6.07)};
                    \addplot[ybar, pattern=north west lines, pattern color=red] coordinates {  (6, 8.60)};
                    \addplot[ybar, pattern=crosshatch, pattern color=cyan] coordinates {  (7, 3.92)};
                    \addplot[ybar, pattern=crosshatch dots, pattern color=cyan] coordinates {  (8, 4.80)};
                    \addplot[ybar, pattern=bricks, pattern color=cyan] coordinates {  (9, 3.64)};
                    \addplot[ybar, pattern=sixpointed stars, pattern color=cyan] coordinates {  (10, 4.32)};
                    \addplot[ybar, pattern=fivepointed stars, pattern color=cyan] coordinates {  (11, 5.80)};

                \nextgroupplot[title={\scriptsize (b) Percentage of public fields to total fields}, 
                xticklabels=\empty]
                    \addplot[ybar, pattern=horizontal lines, pattern color=red] coordinates {  (1, 29.17)};
                    \addplot[ybar, pattern=vertical lines, pattern color=red] coordinates { (2, 34.25)};
                    \addplot[ybar, pattern=grid, pattern color=red] coordinates {  (3, 30.13)};
                    \addplot[ybar, pattern=dots, pattern color=red] coordinates {  (4, 33.66)};
                    \addplot[ybar, pattern=north east lines, pattern color=red] coordinates {  (5, 17.89)};
                    \addplot[ybar, pattern=north west lines, pattern color=red] coordinates {  (6, 12.98)};
                    \addplot[ybar, pattern=crosshatch, pattern color=cyan] coordinates {  (7, 6.18)};
                    \addplot[ybar, pattern=crosshatch dots, pattern color=cyan] coordinates {  (8, 6.32)};
                    \addplot[ybar, pattern=bricks, pattern color=cyan] coordinates {  (9, 9.18)};
                    \addplot[ybar, pattern=sixpointed stars, pattern color=cyan] coordinates {  (10, 7.39)};
                    \addplot[ybar, pattern=fivepointed stars, pattern color=cyan] coordinates {  (11, 9.10)};
                   
                    \nextgroupplot[title={\scriptsize (i) Percentage of public methods to total methods}, xticklabels=\empty]
                     \addplot[ybar, pattern=horizontal lines, pattern color=red] coordinates {  (1, 72.77)};
                    \addplot[ybar, pattern=vertical lines, pattern color=red] coordinates { (2, 75.96)};
                    \addplot[ybar, pattern=grid, pattern color=red] coordinates {  (3, 83.39)};
                    \addplot[ybar, pattern=dots, pattern color=red] coordinates {  (4, 91.4)};
                    \addplot[ybar, pattern=north east lines, pattern color=red] coordinates {  (5, 87.42)};
                    \addplot[ybar, pattern=north west lines, pattern color=red] coordinates {  (6, 86.44)};
                    \addplot[ybar, pattern=crosshatch, pattern color=cyan] coordinates {  (7, 43.88)};
                    \addplot[ybar, pattern=crosshatch dots, pattern color=cyan] coordinates {  (8, 62.82)};
                    \addplot[ybar, pattern=bricks, pattern color=cyan] coordinates {  (9, 48.14)};
                    \addplot[ybar, pattern=sixpointed stars, pattern color=cyan] coordinates {  (10, 46.64)};
                    \addplot[ybar, pattern=fivepointed stars, pattern color=cyan] coordinates {  (11, 46.29)};
          \end{groupplot}
        \end{tikzpicture}
        \ref{CombinedLegendBar}
        \caption{Child classes, public fields and public methods}
        \label{ccpfpm}
    \end{figure*}
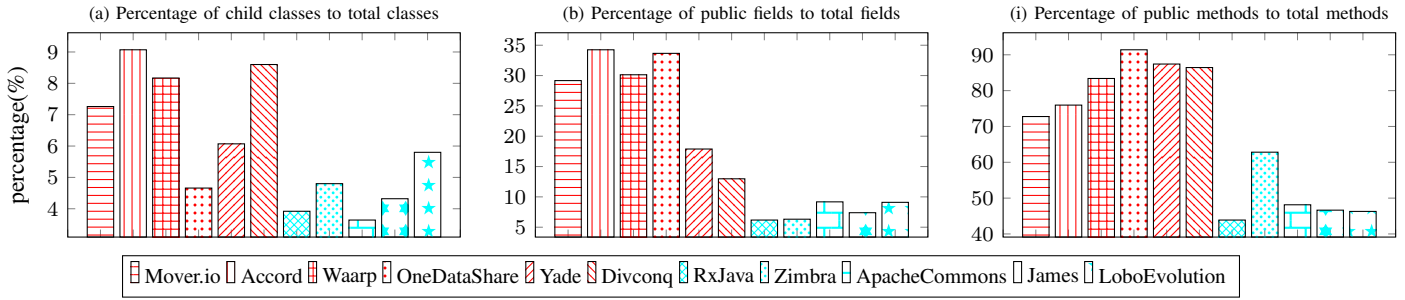
A larger number of classes with respect to the total number of classes is a property when a program uses excessive inheritance operations randomly from existing classes, without considering whether the inheritance is required or not \cite{briand2006toward}. Figure \ref{ccpfpm}a shows the graph of child classes for each of the tested software with respect to the total number of classes in those. Inheritance is used for polymorphism and code re-use. Software which aim to achieve both the aforementioned attributes via single mechanism suffer from inheritance abuse. 

Figure \ref{ccpfpm}a shows that although the proportion of child classes are relatively smaller in established software stack since the long lasting software tend to use composition over inheritance, it is a matter of concern in the stack of new software. Specifically, we see that \textit{Divconq, Waarp, Accord,} and \textit{Yade} have high proportion of child classes. Useful indication for the development team of those software to look into cases can be obtained. We see that the established software stack has a highest proportion of 5.80\% whereas \textit{Divconq, Waarp, OneDataShare, Accord, Mover.io,} and \textit{Yade} have 8.60\%, 8.17\%, 4.66\% 9.07\%, 7.26\% and 6.07\% as shown in Figure \ref{ccpfpm}a,  all higher than the desirable threshold. 

This reflects that the developing software list analyzed here needs to consider the case of large volume of child classes, which, if persisted will lead to the \textit{Wide Hierarchy} design smell which will become uncontrollable in the future. One notable observation is that $OneDataShare$ \cite{imran2018onedatashare}, being a developing software has a lower value of child classes proportionate to total classes. Analysis of the classes of $OneDataShare$ show the developers used more composition than inheritance. 

\subsubsection{Cyclomatic complexity}
\begin{figure}
\resizebox{\columnwidth}{!}{
\begin{tikzpicture}[yscale=0.8]
\begin{axis}[
    ybar,
    enlargelimits=0.10,
    legend style={at={(0.5,0.97)},
      anchor=north,legend columns=-1},
    ylabel={Number of observations (units)},
    xlabel={Analyzed software (name)},
    bar width=6.8pt,
    xtick=data,
    nodes near coords,
    x tick label style={rotate=45,anchor=east},
    symbolic x coords = {Mover.io, Accord, Waarp, OneDataShare, Yade, Divconq, RxJava, Zimbra, ApacheCommons, James, LoboEvolution},
    nodes near coords align={vertical},
    ]

\addplot[pattern=grid, pattern color=orange] coordinates {     (Mover.io,18) (Accord,24) (Waarp,22)  (OneDataShare,13) (Yade,30) (Divconq,33) (RxJava,40) (Zimbra,28) (ApacheCommons,33) (James,44) (LoboEvolution,20)};
  \addplot[pattern=dots, pattern color=black] coordinates {     (Mover.io,29)         (Accord,15) (Waarp,34)  (OneDataShare,5) (Yade,19) (Divconq,17) (RxJava,29) (Zimbra,35) (ApacheCommons,26) (James,10) (LoboEvolution,18)};
  \addplot[pattern=crosshatch, pattern color=green] coordinates {     (Mover.io,19) (Accord,10) (Waarp,26)  (OneDataShare,10) (Yade,21) (Divconq,20) (RxJava,36) (Zimbra,38) (ApacheCommons,37) (James,42) (LoboEvolution,40)};
  
  \draw [thick,red] ({axis cs:Divconq,2} |- {rel axis cs:0,-0.2}) -- 
                             ({axis cs:Divconq,0.44} |- {rel axis cs:0,1.09});

\legend{0<=CC<=19,20<=CC<=49,CC>40}
\end{axis}
\end{tikzpicture}}
\caption{Cyclomatic Complexities (CC) of the analyzed software}
\label{cyclo}
\end{figure}
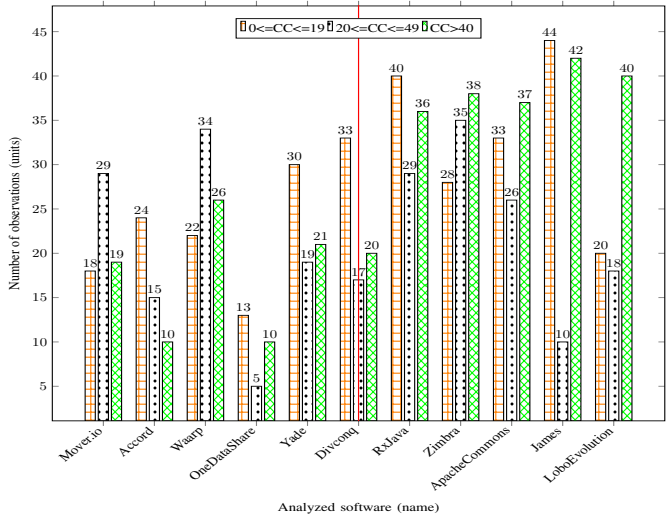

Figure \ref{cyclo} shows the cyclomatic complexity of the new and established software stacks. \textit{ApacheCommons} have the highest cyclomatic complexity among the software in the established stack whereas \textit{Divconq} has the highest cyclomatic complexity among the new software stack. Based on the specifications in \cite{cyclomaticcomplexity}, cyclomatic complexity values between \textit{1 - 19} is considered to be sustainable code, values between \textit{20-39} indicate complex code, \textit{40} and above are considered to have a large number of potentially execution paths and are considered unmaintainable. The red solid line in Figure \ref{cyclo} overlapping on $Divconq$ shows the partition between developing software (left of the red solid line) and established software (right of the red solid line).

\begin{table}
\renewcommand{\arraystretch}{0.9}
\centering
\begin{tabular}{|l|c|}
 \hline
\makecell { \textbf{Software property} }& \makecell {\textbf{Design Smell}}\\ \hline
\scriptsize Child Class & \scriptsize \makecell {Wide Hierarchy, \\Broken Hierarchy}\\ \hline 
\scriptsize Cyclomatic Complexity & \scriptsize \makecell{Insufficient Modularization, \\Cyclic-Dependent Modularization}\\ \hline
\scriptsize Depth of Inheritance Tree & \scriptsize \makecell {Cyclic-Dependent Modularization}\\ \hline
\scriptsize Public Fields & \scriptsize \makecell {Deficient Encapsulation}\\ \hline
\scriptsize Public Methods & \scriptsize \makecell{Deficient Encapsulation, \\Wide Hierarchy}\\ \hline
 \end{tabular}
 \caption{Observed affects of software design properties and occurrence of design smells}
\label{relate}
\end{table}

As seen in Figure \ref{cyclo}, the percentage of program modules with cyclomatic complexity of over 40 and above for the new stack are found to be 3.13\%, 6.76\%, 7.22\%, 13.68\%, 8.64\% and 14.69\% for \textit{OneDataShare, Accord, Mover.io, Waarp, Yade}, and \textit{Divconq} respectively. For the established stack, it is found to be 40.00\%, 42.19\%, 39.80\%, 38.03\% and 36.72\% respectively for \textit{LoboEvolution, JAMES, Zimbra, ApaceCommons, and RxJAVA} respectively. This high value of cyclomatic complexity plays a role in causing \textit{Cyclic-Dependent Modularization} design smell.

\subsubsection{Depth of Inheritance Tree (DIT)}
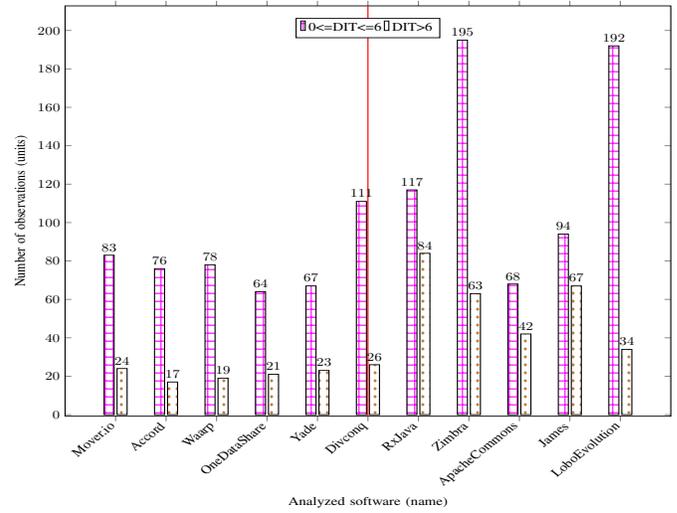
\begin{figure}
\resizebox{\columnwidth}{!}{
\begin{tikzpicture}[yscale=0.8]
\begin{axis}[
    ybar,
    enlargelimits=0.10,
    legend style={at={(0.5,0.97)},
      anchor=north,legend columns=-1},
    ylabel={Number of observations (units)},
    xlabel={Analyzed software (name)},
    bar width=6.8pt,
    xtick=data,
    nodes near coords,
    x tick label style={rotate=45,anchor=east},
    symbolic x coords = {Mover.io, Accord, Waarp, OneDataShare, Yade, Divconq, RxJava, Zimbra, ApacheCommons, James, LoboEvolution},
    nodes near coords align={vertical},
    ]

\addplot[pattern=grid, pattern color=magenta] coordinates {     (Mover.io,83) (Accord,76) (Waarp,78)  (OneDataShare,64) (Yade,67) (Divconq,111) (RxJava,117) (Zimbra,195) (ApacheCommons,68) (James,94) (LoboEvolution,192)};
  \addplot[pattern=dots, pattern color=brown] coordinates {     (Mover.io,24)         (Accord,17) (Waarp,19)  (OneDataShare,21) (Yade,23) (Divconq,26) (RxJava,84) (Zimbra,63) (ApacheCommons,42) (James,67) (LoboEvolution,34)};
  
  \draw [thick,red] ({axis cs:Divconq,2} |- {rel axis cs:0,-0.2}) -- 
                             ({axis cs:Divconq,0.44} |- {rel axis cs:0,1.09});

\legend{0<=DIT<=6,DIT>6}
\end{axis}
\end{tikzpicture}}
\caption{Depth of Inheritance Trees (DIT) of the analyzed software}
\label{ditnew}
\end{figure}

According to Suryanarayana et al. \cite{suryanarayana2014refactoring}, a \textit{DIT} value of six and above signifies a deep hierarchical design which is difficult to understand, causing design smells like \textit{Cyclic-dependent Modularization}. Established software, due to the complicated structure of those, suffer from this smell. Figure \ref{ditnew} shows the results of $DIT$ observations for the analyzed set of software. By focusing on the values of the $DIT$, it is seen that the established software stack have higher number of modules with $DIT$ values greater than 5 such as \textit{RxJava (84), Zimbra (122), ApacheCommons (44), James (74)}, and \textit{LoboEvolution (60}).

\renewcommand{\arraystretch}{0.60}
\setlength{\tabcolsep}{1pt}
\newcommand\Tstrut{\rule{0pt}{2.6ex}}       
\newcommand\Bstrut{\rule[-0.9ex]{0pt}{0pt}} 
\newcommand{\TBstrut}{\Tstrut\Bstrut} 
\begin{table*}
\centering
\scalebox{1.4}{
\scriptsize
\begin{tabular}{|p{4cm}|p{0.6cm}|p{0.6cm}|p{0.6cm}|p{0.6cm}|p{0.6cm}|p{0.6cm}|p{0.6cm}|p{0.6cm}|p{0.6cm}|p{0.6cm}|p{0.6cm}|}
\cline{2-12}
 \multicolumn{1}{c|}{} & \multicolumn{6}{c}{Developing Software} & \multicolumn{5}{|c|}{Established Software}\\ 
\hline
{Design smells} &
 \rotatebox[origin=c]{90}{Mover.io} & \rotatebox[origin=c]{90}{Accord} & \rotatebox[origin=c]{90}{Waarp} & \rotatebox[origin=c]{90}{OneDataShare} & \rotatebox[origin=c]{90}{Yade} & \rotatebox[origin=c]{90}{Divconq} & \rotatebox[origin=c]{90}{RxJava} & \rotatebox[origin=c]{90}{Zimbra} & \rotatebox[origin=c]{90}{ApacheCommons} & \rotatebox[origin=c]{90}{James} & \rotatebox[origin=c]{90}{LoboEvolution} \\ \hline
        Unutilized Abstraction & 164 & 69 & 96 & 56 & 99 & 341 & 884 & 104 & 114 & 56 & 89 \\ \hline
        Insufficient Modularization    & 7 & 9 & 28 & 3 & 22 & 38 & 189 & 62 & 48 & 41 & 43 \\ \hline
        Broken Hierarchy & 49 & 31 & 24 & 16 & 21 & 41 & 47 & 6 & 4 & 0 & 0 \\ \hline
        Deficient Encapsulation     & 39 & 33 & 49 & 19 & 24 & 62 & 18 & 21 & 29 & 16 & 27 \\ \hline
        Cyclic-Dependent Modularization   & 23 & 0 & 17 & 2 & 5 & 102 & 252 & 34 & 30  & 27  & 33 \\ \hline
        Unnecessary Abstraction   & 4 & 3 & 5 & 8 & 0 & 26 & 76 & 15 & 11  & 14  & 12 \\ \hline
        Multifaceted Abstraction   & 3 & 1 & 3 & 1 & 2 & 1 & 18 & 20 & 13  & 11  & 17 \\ \hline
        Wide Hierarchy   & 3 & 3 & 7 & 2 & 6 & 4 & 3 & 1 & 0  & 0  & 0 \\ \hline
        Missing Hierarchy   & 0 & 0 & 1 & 0 & 1 & 3 & 0 & 2 & 0  & 0  & 0 \\
  \hline
    \end{tabular}
    }
\caption{List of top design smells observed for both developing and established software stacks}
\label{fig:top20list}
\end{table*}

\subsubsection{Public fields}
High number of public fields to a class pose a threat by causing $Deficient\ Encapsulation$ design smell. Making a field public will make it accessible by anyone. A software engineer may mistakenly assign a wrong value to the field or maliciously introduce a bug to the public code \cite{yoshioka2008survey}. Thereby, this smell will make the code vulnerable and make the software a target of malicious activities. We see in Figure \ref{ccpfpm}b that the new software stack have significantly higher number of public fields with respect to the number of classes. Opposite behavior is seen for the established software stack which have significantly lower number of public fields with regard to the number of classes. 

Upon analyzing the source codes of the software in established stack, we see those increasingly used private fields and were required to implement a \textit{getter} method to return a clone of the field. The developing stack require this property to be added. So through the mechanism described in this paper, any developer of new software can check the percentage of public fields in their code and take necessary precautions to reduce those at an early stage. This paper compares those properties to make a new developer aware of the threat of having public fields and also allows them to identify the percentage of public fields in their code. In Figure \ref{ccpfpm}b, undesirably high percentage of public fields of 12.98\%, 17.89\%, 29.17\%, 30.13\%, 34.25\% and 33.36\% are seen \textit{Divconq, Yade, Mover.io, Waarp, Accord}, and \textit{OneDataShare}.
\subsubsection{Public methods}
In Figure \ref{ccpfpm}c we present our analysis for the number of public methods normalized by the total number of methods for both the new and established software stacks. Declaring more methods as public increase the surface area of the software which are accessible by everyone. At the same time, if not properly documented, changing behavior of public methods becomes a challenge. Standard software engineering requires us not to change a method to public as long as it is not required. As seen in the figure, despite having comparatively lower number of total methods than the established software stack, the new stack has greater percentage of public methods. 

\textit{Waarp, Divconq, Yade} and \textit{OneDataShare} have high percentage of public methods which are 83.39\%, 86.54\%, 87.42\% and 91.5\% respectively. Despite having greater number of total methods, \textit{RxJAVA, ApacheCommons, LoboEvolution, and JAMES} have percentage values below 50\% (43.88\%, 48.14\%, 46.29\%, and 47.64\% respectively) which is desirable and prevents those from suffering from large volume of $Wide$ $Hierarchy$ as well as $Deficient\ Encapsulation$ \cite{suryanarayana2014refactoring}. A large number of established software have a percentage value of public methods well below 50\% which is desirable and the new software stack codes should be corrected. Table \ref{relate} provides a summary of the causal chain between specific software properties and design smells.

\subsection{Comparison of Design Smells}
Table \ref{fig:top20list} shows the number of observations for nine types of design smells which are detected in the case of developing and established software stacks. Percentage of a smell is calculated by normalizing the number of detection of a specific smell to the total number of smells detected for that software.

Table \ref{fig:top20list} identify which of the design smells occur during both developing and established phases, hence should be given greater importance as they decide to build a new software or maintain a established one. Figure \ref{PlusPlusCombinedBar} provides the percentage of occurrence of each of the observed design smells. We discuss the findings  related to each of the design smells provided in Figure \ref{PlusPlusCombinedBar} in the following.

\subsubsection{Unutilized Abstraction} 
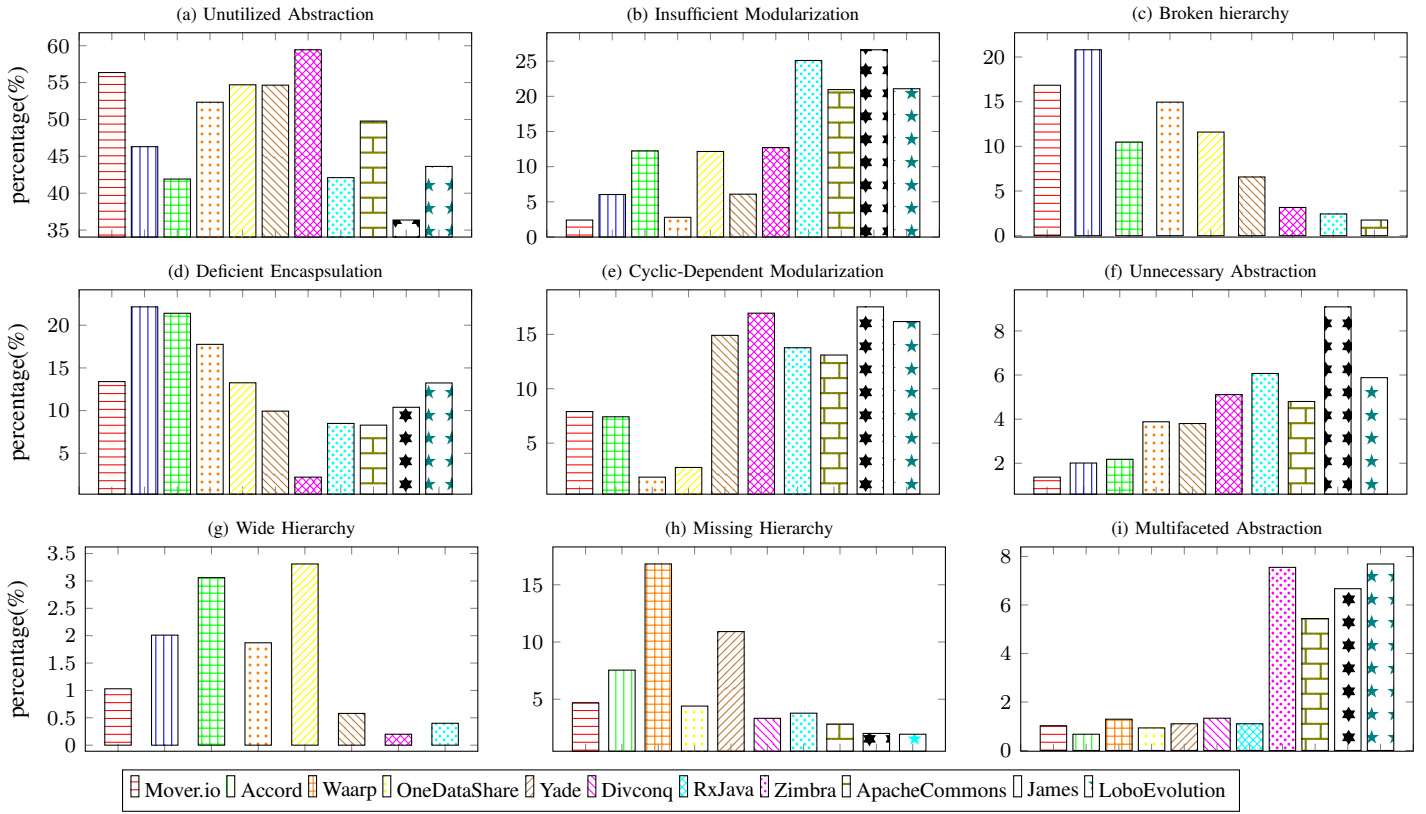
\begin{figure*}
        \centering
        \begin{tikzpicture}
            \begin{groupplot}[
                    legend columns=-1,
                    legend entries={{Mover.io},{Accord},{Waarp},{OneDataShare},{Yade},{Divconq},{RxJava},{Zimbra},{ApacheCommons},{James},{LoboEvolution}},
                    legend to name=CombinedLegendBar,
                    footnotesize,
                    ybar legend,
                    width = 6.8cm,
                    height = 4.3cm,
                    ylabel=percentage(\%),
                    xlabel style={yshift=-1cm},
                    group style={
                    group size=3 by 1,
                    xlabels at=edge bottom,
                    ylabels at=edge left,
                    }]
                \nextgroupplot[title={\scriptsize (a) Unutilized Abstraction}, xticklabels=\empty]
                    \addplot[ybar, pattern=horizontal lines, pattern color=red] coordinates {  (1, 56.36)};
                    \addplot[ybar, pattern=vertical lines, pattern color=blue] coordinates { (2, 46.31)};
                    \addplot[ybar, pattern=grid, pattern color=green] coordinates {  (3, 41.92)};
                    \addplot[ybar, pattern=dots, pattern color=orange] coordinates {  (4, 52.34)};
                    \addplot[ybar, pattern=north east lines, pattern color=yellow] coordinates {  (5, 54.70)};
                    \addplot[ybar, pattern=north west lines, pattern color=brown] coordinates {  (6, 54.65)};
                    \addplot[ybar, pattern=crosshatch, pattern color=magenta] coordinates {  (7, 59.45)};
                    \addplot[ybar, pattern=crosshatch dots, pattern color=cyan] coordinates {  (8, 42.11)};
                    \addplot[ybar, pattern=bricks, pattern color=olive] coordinates {  (9, 49.78)};
                    \addplot[ybar, pattern=sixpointed stars, pattern color=black] coordinates {  (10, 36.36)};
                    \addplot[ybar, pattern=fivepointed stars, pattern color=teal] coordinates {  (11, 43.63)};

                \nextgroupplot[title={\scriptsize (b) Insufficient Modularization}, 
                xticklabels=\empty]
                    \addplot[ybar, pattern=horizontal lines, pattern color=red] coordinates {  (1, 2.41)};
                    \addplot[ybar, pattern=vertical lines, pattern color=blue] coordinates { (2, 6.04)};
                    \addplot[ybar, pattern=grid, pattern color=green] coordinates {  (3, 12.23)};
                    \addplot[ybar, pattern=dots, pattern color=orange] coordinates {  (4, 2.8)};
                    \addplot[ybar, pattern=north east lines, pattern color=yellow] coordinates {  (5, 12.16)};
                    \addplot[ybar, pattern=north west lines, pattern color=brown] coordinates {  (6, 6.09)};
                    \addplot[ybar, pattern=crosshatch, pattern color=magenta] coordinates {  (7, 12.71)};
                    \addplot[ybar, pattern=crosshatch dots, pattern color=cyan] coordinates {  (8, 25.10)};
                    \addplot[ybar, pattern=bricks, pattern color=olive] coordinates {  (9, 20.96)};
                    \addplot[ybar, pattern=sixpointed stars, pattern color=black] coordinates {  (10, 26.62)};
                    \addplot[ybar, pattern=fivepointed stars, pattern color=teal] coordinates {  (11, 21.08)};
                \nextgroupplot[title={\scriptsize (c) Broken hierarchy}, xticklabels=\empty]
                    \addplot[ybar, pattern=horizontal lines, pattern color=red] coordinates {  (1, 16.84)};
                    \addplot[ybar, pattern=vertical lines, pattern color=blue] coordinates { (2, 20.81)};
                    \addplot[ybar, pattern=grid, pattern color=green] coordinates {  (3, 10.48)};
                    \addplot[ybar, pattern=dots, pattern color=orange] coordinates {  (4, 14.95)};
                    \addplot[ybar, pattern=north east lines, pattern color=yellow] coordinates {  (5, 11.60)};
                    \addplot[ybar, pattern=north west lines, pattern color=brown] coordinates {  (6, 6.57)};
                    \addplot[ybar, pattern=crosshatch, pattern color=magenta] coordinates {  (7, 3.16)};
                    \addplot[ybar, pattern=crosshatch dots, pattern color=cyan] coordinates {  (8, 2.43)};
                    \addplot[ybar, pattern=bricks, pattern color=olive] coordinates {  (9, 1.75)};

            \end{groupplot}
        \end{tikzpicture}
        \begin{tikzpicture}
            \begin{groupplot}[
                    legend columns=-1,
                    legend to name=CombinedLegendBar,
                    footnotesize,
                    ybar legend,
                    width = 6.8cm,
                    height = 4.3cm,
                    ylabel=percentage(\%),
                    xlabel style={yshift=-1cm},
                    group style={
                    group size=3 by 1,
                    xlabels at=edge bottom,
                    ylabels at=edge left
                    }]
                \nextgroupplot[title={\scriptsize (d) Deficient Encaspsulation}, xticklabels=\empty]
                    \addplot[ybar, pattern=horizontal lines, pattern color=red] coordinates {  (1, 13.40)};
                    \addplot[ybar, pattern=vertical lines, pattern color=blue] coordinates { (2, 22.15)};
                    \addplot[ybar, pattern=grid, pattern color=green] coordinates {  (3, 21.40)};
                    \addplot[ybar, pattern=dots, pattern color=orange] coordinates {  (4, 17.76)};
                    \addplot[ybar, pattern=north east lines, pattern color=yellow] coordinates {  (5, 13.26)};
                    \addplot[ybar, pattern=north west lines, pattern color=brown] coordinates {  (6, 9.94)};
                    \addplot[ybar, pattern=crosshatch, pattern color=magenta] coordinates {  (7, 2.21)};
                    \addplot[ybar, pattern=crosshatch dots, pattern color=cyan] coordinates {  (8, 8.5)};
                    \addplot[ybar, pattern=bricks, pattern color=olive] coordinates {  (9, 8.30)};
                    \addplot[ybar, pattern=sixpointed stars, pattern color=black] coordinates {  (10, 10.39)};
                    \addplot[ybar, pattern=fivepointed stars, pattern color=teal] coordinates {  (11, 13.24)};

                \nextgroupplot[title={\scriptsize (e) Cyclic-Dependent Modularization}, xticklabels=\empty]
                    \addplot[ybar, pattern=horizontal lines, pattern color=red] coordinates {  (1, 7.90)};
                    \addplot[ybar, pattern=grid, pattern color=green] coordinates {  (2, 7.42)};
                    \addplot[ybar, pattern=dots, pattern color=orange] coordinates {  (3, 1.87)};
                    \addplot[ybar, pattern=north east lines, pattern color=yellow] coordinates {  (4, 2.76)};
                    \addplot[ybar, pattern=north west lines, pattern color=brown] coordinates {  (5, 14.91)};
                    \addplot[ybar, pattern=crosshatch, pattern color=magenta] coordinates {  (6, 16.95)};
                    \addplot[ybar, pattern=crosshatch dots, pattern color=cyan] coordinates {  (7, 13.77)};
                    \addplot[ybar, pattern=bricks, pattern color=olive] coordinates {  (8, 13.1)};
                    \addplot[ybar, pattern=sixpointed stars, pattern color=black] coordinates {  (9, 17.53)};
                    \addplot[ybar, pattern=fivepointed stars, pattern color=teal] coordinates {  (10, 16.18)};

                \nextgroupplot[title={\scriptsize (f) Unnecessary Abstraction}, xticklabels=\empty]
                
                    \addplot[ybar, pattern=horizontal lines, pattern color=red] coordinates {  (1, 1.37)};
                    \addplot[ybar, pattern=vertical lines, pattern color=blue] coordinates { (2, 2.01)};
                    \addplot[ybar, pattern=grid, pattern color=green] coordinates {  (3, 2.18)};
                    \addplot[ybar, pattern=dots, pattern color=orange] coordinates {  (4, 3.88)};
                    \addplot[ybar, pattern=north west lines, pattern color=brown] coordinates {  (5, 3.8)};
                    \addplot[ybar, pattern=crosshatch, pattern color=magenta] coordinates {  (6, 5.11)};
                    \addplot[ybar, pattern=crosshatch dots, pattern color=cyan] coordinates {  (7, 6.07)};
                    \addplot[ybar, pattern=bricks, pattern color=olive] coordinates {  (8, 4.8)};
                    \addplot[ybar, pattern=sixpointed stars, pattern color=black] coordinates {  (9, 9.09)};
                    \addplot[ybar, pattern=fivepointed stars, pattern color=teal] coordinates {  (10, 5.88)};
                   
            \end{groupplot}
        \end{tikzpicture}
                \begin{tikzpicture}
            \begin{groupplot}[
                    legend columns=-1,
                    legend entries={{Mover.io},{Accord},{Waarp},{OneDataShare},{Yade},{Divconq},{RxJava},{Zimbra},{ApacheCommons},{James},{LoboEvolution}},
                    legend to name=CombinedLegendBar,
                    footnotesize,
                    ybar legend,
                    width = 6.8cm,
                    height = 4.3cm,
                    ylabel=percentage(\%),
                    xlabel style={yshift=-1cm},
                    group style={
                    group size=3 by 1,
                    xlabels at=edge bottom,
                    ylabels at=edge left,
                    }]
                \nextgroupplot[title={\scriptsize (g) Wide Hierarchy}, xticklabels=\empty]
                    \addplot[ybar, pattern=horizontal lines, pattern color=red] coordinates {  (1, 1.03)};
                    \addplot[ybar, pattern=vertical lines, pattern color=blue] coordinates { (2, 2.01)};
                    \addplot[ybar, pattern=grid, pattern color=green] coordinates {  (3, 3.06)};
                    \addplot[ybar, pattern=dots, pattern color=orange] coordinates {  (4, 1.87)};
                    \addplot[ybar, pattern=north east lines, pattern color=yellow] coordinates {  (5, 3.31)};
                    \addplot[ybar, pattern=north west lines, pattern color=brown] coordinates {  (6, 0.58)};
                    \addplot[ybar, pattern=crosshatch, pattern color=magenta] coordinates {  (7, 0.20)};
                    \addplot[ybar, pattern=crosshatch dots, pattern color=cyan] coordinates {  (8, 0.4)};
                \nextgroupplot[title={\scriptsize (h) Missing Hierarchy}, xticklabels=\empty]
                    \addplot[ybar, pattern=horizontal lines, pattern color=red] coordinates {  (1, 4.67)};
                    \addplot[ybar, pattern=vertical lines, pattern color=green] coordinates {  (2, 7.54)};
                    \addplot[ybar, pattern=grid, pattern color=orange] coordinates {  (3, 16.82)};
                    \addplot[ybar, pattern=dots, pattern color=yellow] coordinates {  (4, 4.4)};
                    \addplot[ybar, pattern=north east lines, pattern color=brown] coordinates {  (5, 10.91)};
                    \addplot[ybar, pattern=north west lines, pattern color=magenta] coordinates {  (6, 3.33)};
                    \addplot[ybar, pattern=crosshatch dots, pattern color=cyan] coordinates {  (7, 3.78)};
                    \addplot[ybar, pattern=bricks, pattern color=olive] coordinates {  (8, 2.83)};
                    \addplot[ybar, pattern=sixpointed stars, pattern color=black] coordinates {  (9, 2.01)};
                    \addplot[ybar, pattern=fivepointed stars, pattern color=cyan] coordinates {  (10, 1.94)};

                \nextgroupplot[title={\scriptsize (i) Multifaceted Abstraction}, xticklabels=\empty]
                    \addplot[ybar, pattern=horizontal lines, pattern color=red] coordinates {  (1, 1.02)};
                    \addplot[ybar, pattern=vertical lines, pattern color=green] coordinates {  (2, 0.67)};
                    \addplot[ybar, pattern=grid, pattern color=orange] coordinates {  (3, 1.29)};
                    \addplot[ybar, pattern=dots, pattern color=yellow] coordinates {  (4, 0.93)};
                    \addplot[ybar, pattern=north east lines, pattern color=brown] coordinates {  (5, 1.1)};
                    \addplot[ybar, pattern=north west lines, pattern color=magenta] coordinates {  (6, 1.33)};
                    \addplot[ybar, pattern=crosshatch, pattern color=cyan] coordinates {  (7, 1.1)};
                    \addplot[ybar, pattern=crosshatch dots, pattern color=magenta] coordinates {  (8, 7.55)};
                    \addplot[ybar, pattern=bricks, pattern color=olive] coordinates {  (9, 5.44)};
                     \addplot[ybar, pattern=sixpointed stars, pattern color=black] coordinates {  (10, 6.67)};
                      \addplot[ybar, pattern=fivepointed stars, pattern color=teal] coordinates {  (11, 7.69)};
                   
            \end{groupplot}
        \end{tikzpicture}
        \ref{CombinedLegendBar}
        \caption{Percentage (\%) of occurrence of specific design smells to the total number of smells}
        \label{PlusPlusCombinedBar}
    \end{figure*}

High volumes of this smell is detected in cases of both new and established software as shown in Figure \ref{PlusPlusCombinedBar}a. Analysis of the code shows that the software developers declared numerous instances of abstraction which they did not use in the code. The same tendency is seen in the case of established software too. It is important to refactor this design smell from all the software as unused abstraction may mislead software developers in the future. Also, out of all the software which are analyzed, this design smell is the most prominent. Our results are synchronous with the claim of the authors in \cite{fastdesignsmell} which also discovered large volumes of the same design smell. This finding is important for new software developers as it tells them that they should consider \textit{Unutilized Abstraction} has high percentage of occurrence in software analyzed here. The high volume of this smell is prevalent in all the 11 software we have analyzed, with \textit{James} having the lowest \textit{33.94\%} and \textit{RxJava} having the highest percentage of this smell which is \textit{59.45\%}.

\subsubsection{Insufficient Modularization}  This design smell is related to the cyclomatic complexity of the classes. Hence, we can set up a relationship between Figure \ref{cyclo} and Figure \ref{PlusPlusCombinedBar}b. Classes which have high cyclomatic complexity (\textit{RxJava, Zimbra, ApacheCommons, JAMES}) have a greater tendency to exhibit this design smell. The percentage of this smell is higher in established software compared to the developing software stack. Highest frequency of occurrence is seen in \textit{James} with a avalue of \textit{24.85\%} whereas \textit{Mover.io} has the lowest frequency of \textit{2.38\%}. Hence, the software engineers can get two information. This smell occurs as their software becomes more mature and they add more features, so they can consider it while integrating new features. When they write a new class or modify an old one, they should analyze the cyclomatic complexity as it was done in Figure \ref{cyclo}. If they find large values of cyclomatic complexity in the figure, there is increased possibility of occurrence of this smell. 

\subsubsection{Broken Hierarchy} This smell occurs mainly during the initial stages of software development due to design mistakes between supertype and subtypes. Therefore it is more prevalent in developing software. \textit{Accord} has the highest percentage of \textit{20.81\%} out of all its smells, \textit{OneDataShare} has \textit{14.95\%} as shown in Figure \ref{PlusPlusCombinedBar}c. \textit{Mover.io} has \textit{16.67\%} and \textit{Yade} has \textit{11.7\%}. Among the established software stack, \textit{RxJava} has \textit{3.16\%} which is comparatively lower to the developing software. 

\subsubsection{Deficient Encapsulation} It is seen that the newer software stack have a higher percentage of this smell compared to the established stack. Therefore, it can be seen in Figure \ref{PlusPlusCombinedBar}d that the newer software stack has percentage values of \textit{13.27\%, 22.16\%, 21.12\%, 17.76\%, 13.26\%} and \textit{17.84\%} for \textit{Mover.io, Accord, Waarp, OneDataShare, Yade,} and \textit{Divconq} respectively. It suggests that the software developer need to be aware of this smell at early stages of software development when there is a greater chance of it occurring compared to later stages. It can also be related to Figure \ref{ccpfpm}c which shows that a similarity between many public fields results in a software to suffer from this smell.

\subsubsection{Cyclic Dependent Modularization} As discussed earlier, this smell occurs when the software violates the acyclic modularization technique \cite{sarkar2012measuring}. It occurs when one abstraction is cyclically dependent on another. If the cyclic relationship is complicated, it is a challenge to detect that smell. As seen in Figure \ref{PlusPlusCombinedBar}e, the frequency of this smell is more prevalent in established software, which underwent years of modifications and contains complex functionality. 
The cyclic dependencies may be very subtle to detect as it requires more information rather than a rule-based solution. Hence \textit{RxJava, JAMES, LoboEvolution, Zimbra,} and \textit{ApacheCommons} all contain high percentage of this smell which are (\textit{16.96\%, 16.36\%, 14.93\%, 12.83} and \textit{12.66\%}) respectively. The frequency of occurrence of this smell is relatively less in developing software stack. As seen in Figure \ref{PlusPlusCombinedBar}e, \textit{Accord} had no presence of this smell. 

\subsubsection{Unnecessary Abstraction} When a software has multiple layers of abstraction and analysis of software code by engineers determine that those abstraction layers are not required, then the software is said to suffer from unnecessary abstractions \cite{unnecessary}. The best software practice is to eliminate this smell at the design level, since refactoring may become challenging later because the software engineer consider the various levels of abstractions \cite{unnecessary}. Figure \ref{PlusPlusCombinedBar}f identifies the percentage of this design smell for each of the software compared to the total number of observed smells. 

It can be seen that as software graduate from developing stack to established stack, there is a higher percentage of occurrence of this smell. For example, it can be inferred from Figure \ref{PlusPlusCombinedBar}f that \textit{JAMES, Zimbra, LoboEvolution, RxJava, ApacheCommons,} and \textit{Zimbra} have a frequency of occurrence of \textit{8.48\%, 5.66\%, 5.43\%, 5.11\%} and \textit{4.6\%} respectively of this smell compared to the total number of smells in comparison to software in developing stack which have the highest frequency of occurrence value of \textit{3.88\%}. Hence it is clear from Figure \ref{PlusPlusCombinedBar}f that this smell is more prevalent as a software gets established. 

\subsubsection{Wide Hierarchy} This smell is prevalent in all the developing stack and absent in all but 2 of the software in established stack. Therefore, software engineers need to be aware that it might be present in their newly developed code and take the required steps to refactor it. The percentage of this smell is lower compared to the total observed smells, for example, \textit{Mover.io, Accord, Waarp, OneDataShare, Yade,} and \textit{Divconq} have lower percentage values of \textit{1.02\%, 2.01\%, 3.02\%, 1.94\%, 3.31\%} and \textit{0.58\%} respectively. 

\subsubsection{Imperative Abstraction} As stated earlier, if an operation is converted to a class, i.e. the class has limited functionality in an operation, it becomes a case of \textit{Imperative Abstraction}. As expected, this design smell occurs mostly in new software, where it covers \textit{0.69\%, 0.86\%, 0.55\% and 0.88\%} of total number of observed smells in \textit{Mover.io, Waarp, Yade,} and \textit{Divconq}. Our experimental runs revealed that this smell is present only in new software, the software developers need to consider it during development of a new software. The established software have comparatively well-planned implementations of classes and hence do not have this smell.

\subsubsection{Multifaceted Abstraction} This smell is predominant in established software. It occurs when an abstraction has multiple responsibilities. \textit{LoboEvolution} has 7.64\% of frequency of occurrence of this smell out of total number of smells in this category which is highest. It may be said that the developers should be aware of this smell throughout the later stages as a software moves towards maturity. 

To summarize the results, it can be seen that the software engineers should worry about \textit{Insufficient Modularization, Cyclic-Dependent Modularization, Unnecessary Abstraction,} and \textit{Multifaceted Abstraction} smells in the case of established software. \textit{Broken Hierarchy, Deficient Encapsulation, Imperative Abstraction, Wide Hierarchy,} and \textit{Missing Hierarchy} smells tend to occur during developing phase due to undesirable design practices applied at the initial phases of software development. Hence, within the boundary of those design smells, software engineers can focus on the ones which tend to occur on the new software stack. This will save them cost in terms of time and money as they focus only on a subset of design smells and ignore the rest, hence reducing technical debt.

The results analyzed in this section provide useful information to a software engineer aiming to design and develop a new distributed software, allowing them to focus on specific design smells which happen more frequently during early stages of software development life-cycle. Also, it helps the software engineers of established software to identify which design smells occur at the maturity phase of software, thus allowing them to focus on refactoring only those. Therefore, software engineers can deploy this model on their new and old software and highlight the required design smells which are needed to be addressed depending on the age of the software, thus they have to deal with a less number of design smells which saves them important time and money, hence reducing technical debt.

\renewcommand{\arraystretch}{0.90}
\setlength{\tabcolsep}{1pt}
\begin{table*}
\centering
\scalebox{1}{
\begin{tabular}{|p{2.7cm}|p{1cm}|p{1cm}|p{1cm}|p{1cm}|p{1cm}|p{1cm}|p{1cm}|p{1cm}|p{1cm}|}
\hline
 \diaghead{\theadfont DiagColumnmnHead II}%
  {\\~~~Software~}{~Design~~\\ ~Smell~~} & \thead{\rotatebox[origin=c]{90}{\shortstack[4]{\makecell{Unutilized\\ Abstraction}}}} & \thead{\rotatebox[origin=c]{90}{\makecell{Insufficient\\ Modularization}}} & \thead{\rotatebox[origin=c]{90}{\makecell{Broken\\ Hierarchy}}} & \thead{\rotatebox[origin=c]{90}{\makecell{Deficient\\ Encapsulation}}} & \thead{\rotatebox[origin=c]{90}{\makecell{Cyclic\\ Modularization}}} & \thead{\rotatebox[origin=c]{90}{\makecell{Unnecessary\\ Abstraction}}} & \thead{\rotatebox[origin=c]{90}{\makecell{Multifaceted\\ Abstraction}}} & \thead{\rotatebox[origin=c]{90}{\makecell{Wide\\ Hierarchy}}} & \thead{\rotatebox[origin=c]{90}{\makecell{Missing\\ Hierarchy}}} \TBstrut \\ \hline
       \makecell{OneDataShare \\ (Developing software)} & \makecell{ 60 \\ 56\\93.3\%} &  \makecell{4 \\ 3 \\ 75\%} & \makecell{17 \\ 16 \\ 94.1\%} & \makecell{19 \\ 19 \\ 100\%} & \makecell{2 \\ 2 \\ 100\%} & \makecell{8 \\ 8 \\ 100\%} & \makecell{2 \\ 1 \\ 50\%} & \makecell{3 \\ 2 \\ 66.7\%} & \makecell{2 \\ 1 \\ 50\%} \\ \hline
         \makecell{Accord\\ (Developing software)}  & \makecell{81 \\ 69 \\ 85.2\%} & \makecell{13 \\ 9 \\ 69.2\%} & \makecell{39 \\ 31 \\ 79.5\%} & \makecell{42 \\ 33 \\ 78.6\%} & \makecell{0 \\ 0 \\ 100\%} & \makecell{4 \\ 3 \\ 75\%} & \makecell{2 \\ 1 \\ 50\%} & \makecell{3 \\ 3 \\ 100\%} & \makecell{0 \\ 0\\ 100\%}  \\ \hline
       \makecell{James\\ (Established software)} & \makecell{70 \\ 56 \\ 80.01\%} & \makecell{46 \\ 41 \\ 89.1\%} & \makecell{0 \\ 0 \\ 100\%} & \makecell{16 \\ 16 \\ 100\%} & \makecell{30 \\ 27 \\ 90\%} & \makecell{16 \\ 14 \\ 87.49\%} & \makecell{18 \\ 11 \\ 61.1\%} & \makecell{0 \\ 0 \\ 100\%} & \makecell{0 \\ 0 \\ 100\%} \\ \hline
         \makecell{LoboEvolution \\ (Established software)}  & \makecell{103 \\ 89 \\ 86.4\%} & \makecell{47 \\ 43 \\ 91.5\%} & \makecell{0 \\ 0 \\ 100\%} & \makecell{33 \\ 27 \\ 81.8\%} & \makecell{38 \\ 33 \\ 86.8\%} & \makecell{12 \\ 12 \\ 100\%} & \makecell{18 \\ 17 \\ 94.4\%} & \makecell{0 \\ 0 \\ 100\%} & \makecell{0 \\ 0 \\ 100\%}  \\ \hline
    \end{tabular}
    }
    \caption{Results of the design smell determination mechanism for nine types of smells. \textit{(In each row, the first line identifies the quantity of number of suspected smells detected by the tool, the second line shows the number of actual design smells (true positives), the third line shows the precision as percentage.)} }
\label{fig:precision}
 \end{table*}   
 
\subsection{Precision and Recall of the Detection Mechanism}

In this subsection we analyze the precision and recall for the design smell detection tool. Previous results are capable of detecting four types of design smells for Java \cite{moha2009decor} whereas this approach detects nine types of smells with desirable accuracy. 

Suspicious classes were identified during the detection phase and those were manually analyzed by a team to validate the findings. True positives were determined and the process validated. The tool detected a total of 4,020 smells by analyzing 17,760 Java classes for developing and established source codes combined. We extend the effort to the area of information retrieval and detect precision and recall \cite{frakes1992information}. More specifically, precision identifies the smells out of the total which could be successfully detected. Recall assesses the total number of detected and undetected smells. 

Table \ref{fig:precision} present the precision and recall values of 2 developing and 2 developed software from the list of 11 software analyzed during this research. 

For developing software, overall precision value for 9 smells were obtained as \textit{72.9\%} for \textit{OneDataShare}. For \textit{Accord}, precision values ranged from \textit{50.0\%} till \textit{100.0\%} for the specific smells, with an overall precision of \textit{73.8\%}. The low percentage of classes which were suspected allowed manual analysis of those within a reasonable time frame compared to having to analyze a total of 266 classes otherwise for those two software only. The results can be considered reflective to all the other remaining developing software.

Similar analysis were done for \textit{James} and \textit{LoboEvolution} from the list of established software analyzed here. The overall precision for \textit{James} and \textit{LoboEvolution} were found to be \textit{80.8\%} and \textit{84.1\%} respectively. The detection approach has a overall precision of over 70\% in all the analyzed cases, and outperforms the technique specified in \cite{moha2009decor}. Additionally, the obtained results are significantly better than random chance. The algorithms has a \textit{100\%} recall value since no smell were missed within the scope of the nine types of smells considered in this research. Hence, the results obtained can be generalized to the wide array of developing and established software.  

The affect of the observed design smells on new and established software stacks based on the software tested in this research is shown in Table \ref{fig:summary}. The table stated the design smells considered in this paper, next it focused on identifying the overall precision for developing and established software. Hence, it shows the performance of the smell detection tool. 
  
\begin{table}[]
\begin{center}
    \centering
    \begin{tabular}{|p{4cm}|P{2cm}|P{2cm}|}
    \hline
\textbf{Design Smell }             & \textbf{Impact:Dev(\%)} & \textbf{Impact:Estb(\%)}\\
\hline
Unutilized Abstraction & 87.16 & 83.3   \\
\hline
Insufficient Modularization           & 72.1 & 90.3  \\
\hline
Broken Hierarchy           & 86.8 & 100    \\
\hline
Deficient Encapsulation           & 89.3 & 90.9   \\
\hline
Cyclic Modularization           & 100 & 88.4    \\
\hline
Unnecessary Abstraction          &  87.5& 93.7   \\
\hline
Multifaceted Abstraction           & 49.4 & 77.8    \\
\hline
Wide Hierarchy     & 83.4 & 100    \\
\hline
Missing Hierarchy     & 75.0 & 100   \\
\hline

\end{tabular}
    \caption{Cumulative analysis of impact of the observed design smells}
    \label{fig:summary}
    \end{center}
\end{table}

\subsection{Implications of Results}
The analyses provided in this paper can be applied to software companies, software engineers and researchers.
\begin{itemize}
\item\textbf{Usefulness to software companies:} Firstly, managers in software companies can apply their projects to the tool discussed here, collect the data and analyze results to improve their software engineering processes. Also, by keeping the results of the paper in mind, attention can be paid to only specific smells depending on whether it is a developing or established software.
\item\textbf{Usefulness to software engineers:} The software engineers, both working for companies or as freelancers, can use the results of this paper to identify which kind of design smells are more prevalent based on whether the software they are re-factoring is developing or established. Also they can use the tool discussed here to detect design smells with desirable precision. 
\item\textbf{Usefulness to researchers:} Researchers can use this information to identify which design smells occur specifically during the developing and established stages of software. This will provide them useful lead on which type of smells to focus their research on based on the type of software they are using for experimentation. 
\end{itemize}
By detecting and designating the occurrence of specific smells for developing and established software, this paper allows engineers and researchers to re-factor only a subset of smells at a time.
\section{Related Work}
\label{relatedwork}
A review of exiting tools to detect architectural smells is provided by Azadi et.al. \cite{azadi2019architectural} which groups those based on the detected smells. The authors evaluated 9 tools which are currently in place and ignored those which became obsolete. Smell definitions on which the tools detect the smells are provided and a high level detection mechanism is described by analyzing the current literature. Although the tools which detect architectural smells are provided, how the detected output varies from developing to established software to aide developers regarding which smells to focus on during different stages of software life cycle was not addressed.

Model depicting the collected efforts required to detect architectural design smells is proposed based on study of related literature \cite{besker2016systematic}. The model identifies negative affects, challenges, refactoring areas and relationship between each activity. The design of the model was motivated from the literature study conducted earlier in the paper. Although the model considers important aspects of design smells, it is a high level design and performance analysis of it is not provided as it was not tested on any software. Also the ability of the proposed architecture to detect smells of developing and established software was not discussed.

Technique to prioritize design smells based on design best practices with an aim to reduce technical debt is proposed by Pl̈osch et.al.  \cite{plosch2018design}. Through the application of a bench marking technique, the non-conformance of software systems to address those issues was determined. Next, a portfolio matrix is produced which enabled the stakeholders to decide which design smell should have higher priority. The stability of the proposed approach is illustrated by testing on an open source project. However, the importance of prioritizing design debts for developing and established $OSS$ to enable software developers to decide which smells to prioritize at different stages of software development was not provided.

The affect of developer's seniority, frequency of commits and interval of commits on reducing design debts in a software was evaluated by Alfayez et.al.  \cite{Alfayez:2018:ESI:3194164.3194165}. The authors determined that seniority and frequency of commits are negatively correlated to reducing design debt, whereas interval of commits is positively correlated. The authors used multiple statistical analysis tests to validate the affects of developer behavior on the design smells. However, whether developer behavior differently affects design smells in newly developing and already established software was not analyzed.  

Although many software analysis tools provide information on various metrics, there is a limitation in the number of open source tools which can detect design smells \cite{biaggi2018architectural}. \textit{AI Reviewer} identifies code and design smells for C++ projects \cite{azadi2019architectural}. \textit{Hotspot Detector} is capable of detecting only 4 design smells in Java \cite{mo2015hotspot}. \textit{Designite} is a commercial software which detects 25 design smells for C\# projects only \cite{7809461}. \textit{Lattix} is another commercial tool which uses dependency matrix to detect modularity violations \cite{wong2011detecting}. Other commercial tools such as \textit{Structure101, Sonargraph} and \textit{Cast} detect cyclic dependency smells \cite{roveda2018identifying}. To the best of our knowledge, although there are tools to detect code smells, those which detect design smells for Java are limited to detection of 4 or less smells.

\section{Conclusion and Future Work}
\label{conclusion}
This paper detected design smells using an \textit{Abstract Syntax Tree} ($AST$) based tool. It identifies specific design qualities and sets up relationships between violation of those and occurrence of design smells. After the tool was applied on the large volume of LoC and class files of the Java projects analyzed here, the following conclusions were drawn.

\begin{itemize}
    \item In general there is relationship between specific software quality properties and the occurrence of design smells.
    \item Certain design smells are more prevalent on developing software and other smells are prevalent in established software. 
\end{itemize}

The results indicate that high volume of a smell in developing software are present in minute amount in established software, and vice-versa. Software engineers do not need to focus on the entire list of design smells, they can reduce design debt by concentrating on a specific set of smells based on whether the software is developing or established. Analysis of precision showed values ranging from \textit{72.9\%} to \textit{84.1\%} both for developing and established software stacks which confirmed that the mechanism achieved desirable performance in detecting the smells. Recall was found to be 100\% since all the pre-specified types of design smells could be covered.

In the future, it may also be helpful to perform a longitudinal study that detects how the occurrence of design smells in a program suite changes over time. Another interesting point is that the pseudo-model tool can be extended to include a new layer consisting of non-technical factors which affect the design smell.

\bibliographystyle{IEEEtran}
\bibliography{base}
\vspace{12pt}

\end{document}